\documentclass[floatfix,groupedaddress,amsmath,amssymb,twocolumn]{revtex4}
\usepackage{graphicx} 
\usepackage{epsfig} 
\usepackage{dcolumn}
\usepackage{bm}
\usepackage{color}

\newcommand{\be}{\begin{eqnarray}}
\newcommand{\ee}{\end{eqnarray}}
\newcommand{\nn}{\nonumber}
\def\bt{\bibitem}

\def\beq{\begin{equation}}
\def\eeq{\end{equation}}
\def\beqa{\begin{eqnarray}}
\def\eeqa{\end{eqnarray}}
\def\ban{\begin{eqnarray*}}
\def\ean{\end{eqnarray*}}
\def\bi{\begin{itemize}}
\def\ei{\end{itemize}}

\begin{document}

\title{Nuclear symmetry energy and core-crust transition in neutron stars: a critical study}

\author{Camille Ducoin$^1$, J\'er\^ome Margueron $^2$, Constan\c ca Provid\^encia $^1$}
\affiliation{$^1$ CFC, Department of Physics,
University of Coimbra, P3004 - 516, Coimbra, Portugal\\
$^2$Institut de Physique Nucl\'eaire, Universit\'e Paris-Sud, IN2P3-CNRS, F-91406 Orsay Cedex, France}

\begin{abstract}
The slope of the nuclear symmetry energy at saturation density $L$ is pointed out as a crucial quantity to determine the mass and width of neutron-star crusts.
This letter  clarifies the relation between $L$ and the core-crust transition. We confirm that the transition density is soundly correlated with $L$ despite differences between models, and we propose a clear understanding of this correlation based on a generalised liquid drop model. 
Using a large number of nuclear models, we evaluate the dispersion affecting the correlation between the transition pressure $P_t$ and $L$. 
From a detailed analysis it is shown that this correlation is weak due to a cancellation between different terms. 
The correlation between the isovector coefficients $K_{\rm sym}$ and $L$ plays a crucial role in this discussion.
\end{abstract}

\maketitle

Stimulated by the development of exotic nuclear physics, the efforts to determine the nuclear
equation of state (EOS) have focused in the last few years on the density dependence of the
symmetry energy 
{$S(\rho)$~\cite{iida07,Xu09}}. 
In particular, the symmetry-energy slope at saturation density, represented by the quantity $L$, 
has raised a great deal of interest \cite{typel01,steiner05,iida07,Xu09,centelles09,vidana09}: 
while the different nuclear models widely disagree on the value of this basic quantity, 
increasing experimental data \cite{dan09,li05,famiano06,shetty07,klim07,trippa08,centelles09} are expected to bring more and more stringent constraints, 
leading to a radical progress in our knowledge of the EOS of neutron-rich matter. {Several experimental
constraints on the slope $L$ have already been proposed in the last decade. Such studies include
information obtained from the mass formula
\cite{dan09}, isospin diffusion \cite{li05}, experimental double neutron to proton ratio
\cite{famiano06}, and  isoscaling  parameters in heavy ion collisions
\cite{shetty07},  pygmy dipole resonances \cite{klim07}, giant dipole resonances
\cite{trippa08},  neutron-skin thickness \cite{centelles09}.}
This impacts strongly on the physics of compact stars. 
In this letter, we will discuss the link between $L$ and the transition from the liquid core to the solid crust of a neutron star. 
It has been claimed that a precise determination of $L$ 
would give a tight indication of the density $\rho_t$ and pressure $P_t$ at the transition point \cite{Xu09}, 
and consequently the mass and extension of the crust which play a crucial role 
in the interpretation of pulsar observations \cite{pethick95}. 
However, the role of $L$ in the determination of the core-crust transition needs to be checked against model dependence and clarified,
as mentioned in Ref. \cite{pasta}.
In the present work, we use a variety of  nuclear models to address this issue.
We verify and explain the strong correlation between $L$ and $\rho_t$. 
However, we show that when independent models are considered there is no real correlation between $L$ 
and the pressure at the transition point. 
This behaviour results from a competition between opposite effects which destroy the correlation. 
This serious limitation has to be taken into consideration when drawing astrophysical consequences 
from the experimental determination of $L$.

Catalyzed matter in compact stars satisfies the $\beta$-equilibrium
condition which favors very neutron-rich matter: 
the proton fraction is reduced to a few percent in the region of the core-crust transition. 
Compact-star structure crucially depends on the symmetry energy, for a wide density range. 
The density dependence of the symmetry energy, $S(\rho)$, is 
{deduced from}
the energy density functional obtained in the framework of mean field nuclear models. 
Besides, it can be expressed as a development around the saturation  {density $\rho_0$},
whose coefficients correspond to the isovector parameters of a generalised liquid-drop 
model (GLDM):
\be
S(\rho)&=&\sum_{n\ge 0}c_{{\rm IV}, n}\frac{x^n}{n!}\,,
\ee
where $x=(\rho-\rho_0)/(3\rho_0)$. 
{ Here and in the sequel, the  index "IV" ("IS")  attributed to the
coefficients of the GLDM stands  for  "isovector" ("isoscalar").} The first coefficients have received traditional denominations:
{$c_{{\rm IV}, 0}=J\equiv S(\rho_0)$,}
$c_{{\rm IV}, 1}=L$, 
$c_{{\rm IV}, 2}=K_{\rm sym}$, etc. 
In the framework of the parabolic approximation,
the energy per particle for asymmetric matter is given by
{$E(\rho,y)=E(\rho,0)+{S}(\rho)y^2$, }
where $y=(\rho_n-\rho_p)/\rho$. {For convenience, we will use in the following
either the isospin-asymmetry $y$ or the proton fraction $Y_p=(1-y)/2$.}
This approximation allows to emphasise the role of the GLDM coefficients, so we will use it to
{analyze our results,}
although the calculations have been performed using the complete density functional of each model.
In the parabolic-GLDM framework, the energy per particle  reads:
\be
E(\rho,y)&=&\sum_{n\geq 0}\left(c_{{\rm IS}, n}+c_{{\rm IV}, n}y^2\right)\frac{x^n}{n!}\, .
\ee
{In the isoscalar channel, we have 
$c_{{\rm IS}, 0}=E_0\equiv E(\rho_0)$, $c_{{\rm IS}, 1}=0$, $c_{{\rm IS}, 2}=K_{\infty}$, etc.}

We will show results obtained from a set of non-relativistic and relativistic
effective interactions, together with results from a 
microscopic Brueckner-Hartree-Fock (BHF) calculation using the interaction Av18
\cite{wiringa95} with Urbana three-body forces \cite{baldo99}.
As non-relativistic effective models, 
we take Skyrme-type interactions from different groups
(SV, SGII, RATP, SkMP, Gs, Rs, SkI2, SkI3, SkI4, SkI5, SkI6, Sly10, Sly230a, Sly230b, Sly4, SkO, NRAPR, LNS, BSk14, BSk16, BSk17); 
the respective references can be found in \cite{sky-rel,Stone,Gor17,chabanat97}.
Besides, we consider two different types of relativistic effective nuclear models: 
(i) non-linear Walecka models with constant couplings 
(NL3 \cite{nl3}, TM1 \cite{tm1}, GM1, GM3 \cite{gm91}, FSU, NL$\omega\rho$ \cite{fsu}, NL$\rho\delta$ \cite{liu02}); 
(ii) hadronic models with density dependent coupling constants 
(TW \cite{tw}, DD-ME1, DD-ME2 \cite{ddme}, DDH$\delta$ \cite{ddhd}).
Let us remark that the EOS features present more variability within the relativistic models
than within the Skyrme ones \cite{sky-rel}.

The inner crust of a neutron star is usually modelised as a lattice of very neutron-rich nuclei, 
immersed in a gas of electrons and dripped neutrons. 
As the density increases, the difference between the nuclei and surrounding neutron gas decreases, 
until the stellar matter becomes homogeneous: this is the transition to the liquid core, {when the pasta phase dissolves \cite{maruyama05,iida07,pasta}.} 
In order to determine the transition point, one should in principle compare the free energy of
homogeneous matter to that of any inhomogeneous configuration, {which constitutes the pasta
phase} \cite{maruyama05}. 
{It has been shown in \cite{steiner08} that the composition of the crust, namely the size of
the
pasta structures, is sensitive to the very low density neutron matter EOS \cite{tolos08}
while almost not affecting the crust-core transition.  Therefore, we expect that our
conclusions will not depend on the very low density behavior of the neutron matter EOS.}
In \cite{iida07,pasta}
it has been verified that the transition density obtained from the pasta phase calculation can be very well approximated 
by the entrance into the dynamic spinodal region, under the constraint of $\beta$ equilibrium. 
The spinodal is the density region where the homogeneous matter is unstable against density fluctuations, 
due to the nuclear liquid-gas phase transition affecting the bulk EOS. 
In the case of finite size density fluctuations, the Coulomb and surface terms reduce the instability: 
the dynamic spinodal region is then smaller than the thermodynamic one obtained when only the bulk term is considered. 
The difference between the two regions is in principle model dependent, via the nuclear surface term. 
However, we found that this effect is too small to play a role in the present discussion. 
Thus, for simplicity, we will focus on the transition density $\rho_{t}$, 
proton fraction $Y_{p,t}$ 
and   pressure $P_t$ 
taken at the crossing point between the $\beta$ equilibrium EOS of stellar matter and the thermodynamic spinodal, 
keeping in mind that they represent shifted values of the actual density, 
proton fraction and pressure at the core-crust transition.

\begin{figure}[t]
\begin{center}
\begin{tabular}{ccc}
\includegraphics[width =.9\linewidth]{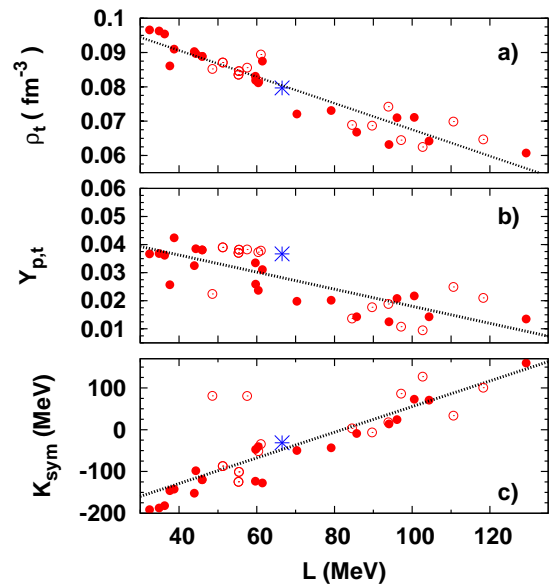}&
\end{tabular}
\end{center}
\caption{(Colour online) 
Correlation between $L$ and 
a) the transition density $\rho_t$, 
b) the  transition proton fraction $Y_{p,t}$, 
c) $K_{\rm sym}$.
The full (empty) symbols are for Skyrme forces (relativistic models) and the asterisk for BHF.
}
\label{Fig:L0_rt_Ypt_Ksym}
\end{figure}

\begin{figure}[t]
\begin{center}
\begin{tabular}{cc}
\includegraphics[width =0.9\linewidth]{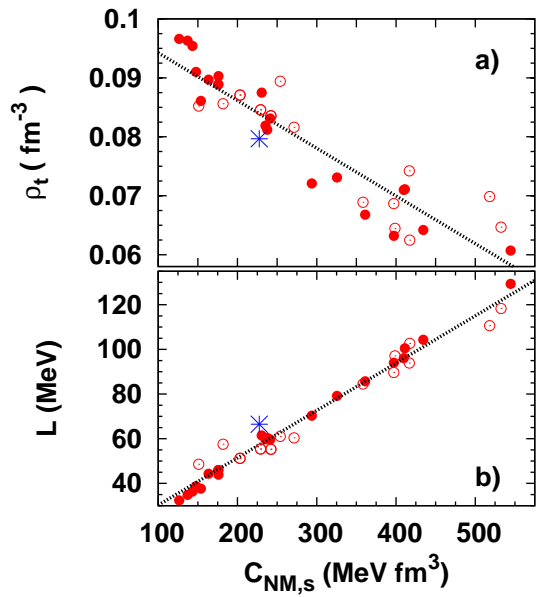}&
\end{tabular}
\end{center}
\caption{(Colour online) 
Correlation of the energy-density curvature of neutron matter $C_{{\rm NM},s}$ with
a) the transition density $\rho_t$, and
b) the symmetry energy slope $L$. 
The full (empty) symbols are for Skyrme forces (relativistic models) and the asterisk for BHF.
}
\label{Fig:Cnms}
\end{figure}

It has been noticed in previous works that the transition density decreases as $L$ increases \cite{iida07,Xu09,vidana09}; 
this correlation has been verified with many different models, see Fig.~\ref{Fig:L0_rt_Ypt_Ksym}a).
We have found that this behaviour can be understood 
through the energy-density curvature of pure neutron matter (NM), denoted $C_{\rm NM}$. 
More specifically, we have considered {the curvature} $C_{\rm NM}$ taken at the density of the upper spinodal border in symmetric matter, $\rho_s$.
This quantity, denoted $C_{{\rm NM},s}=C_{\rm NM}(\rho_s)=\frac{d^2(\rho E_{\rm NM})}{d\rho^2}(\rho_s)$, {where $E_\mathrm{NM}$ is the energy per particle in neutron matter,}
is {indeed} correlated with the transition density $\rho_t$, as shown in Fig.~\ref{Fig:Cnms}a).
This result allows a qualitative interpretation.
The spinodal region corresponds to the region of $(\rho,\, Y_p)$ 
where the energy density has a negative curvature. 
For very asymmetric matter such as $\beta$-equilibrium matter, 
$C_{{\rm NM},s}$ gives a good indication 
{to localise the position of the spinodal border:}
the larger $C_{{\rm NM},s}$ is, 
the farther should be the spinodal contour from the point $(\rho=\rho_s,Y_p=0)$;
and this corresponds to a lower $\rho_t$.
Besides, $C_{{\rm NM},s}$ is strongly correlated with $L$: see Fig~\ref{Fig:Cnms}b).
This relation appears clearly when $C_{{\rm NM},s}$ is expressed in the parabolic approximation, 
in terms of the isovector coefficients $c_{{\rm IV},n}$:
\be
\label{Eq:Cnms}
C_{{\rm NM},s}
&=&
\frac{2}{3\rho_0}L + 
\frac{1}{3\rho_0}\sum_{n\geq 2}
 c_{{\rm IV},n}\nn
\frac{x_{s}^{n-2}}{(n-2)!}
\left[ \frac{n+1}{n-1}x_{s}+\frac{1}{3}  \right]\, .
\ee
Note that the isoscalar terms of this expansion are exactly zero at $\rho=\rho_s$ by definition of $\rho_s$.
Since $x$ is negative, the influence of the higher order terms $n\geq 2$ is weakened.
Furthermore, for all the models considered, we have $\rho_s\simeq (2/3)\rho_0$:
this makes the contribution of the term $n=2$ in Eq.~(\ref{Eq:Cnms}) close to zero.
As a result, $C_{{\rm NM},s}$ depends very weakly on $K_{\rm sym}$ and is essentially determined by $L$,
as can be verified on the figure.
In summary, the correlation observed between $L$ and $\rho_t$ can be understood 
as the consequence of the link existing between {$L$,} $C_{{\rm NM},s}$ and $\rho_t$.
%

The proton fraction at the transition, $Y_{p,t}$, is also expected to decrease with increasing $L$.
Indeed, a smaller symmetry energy corresponds to a lower proton fraction for $\beta$ equilibrium matter. 
Assuming a consensual value of {$J$} (about 32 MeV),  
a larger $L$ means a smaller symmetry energy at subsaturation densities and, consequently, 
a smaller $Y_{p,t}$, as is shown in Fig.~\ref{Fig:L0_rt_Ypt_Ksym}b).  
The dispersion of data in this figure reflects the model dependence of $J$.
We also show in Fig.~\ref{Fig:L0_rt_Ypt_Ksym}c) 
the correlation between $K_\mathrm{sym}$ and $L$, which will be useful for the following analysis.

\begin{figure}[t]
\begin{center}
\begin{tabular}{ccc}
\includegraphics[width =0.9\linewidth]{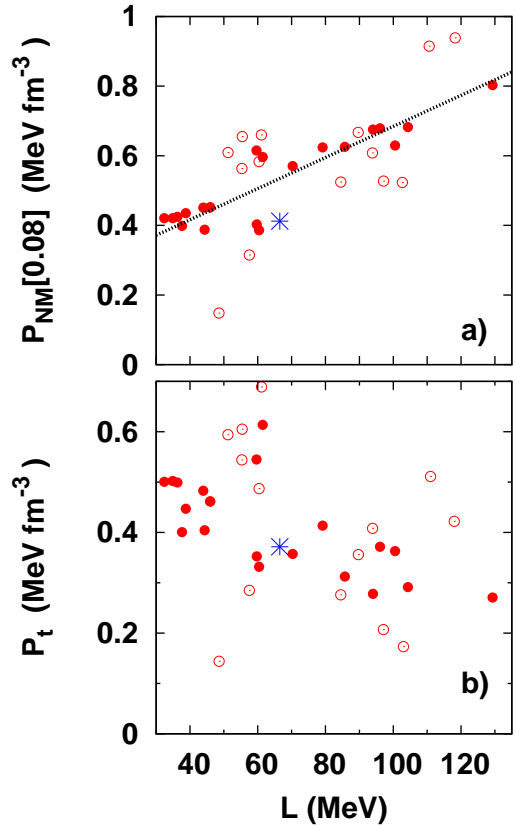}&
\end{tabular}
\end{center}
\caption{(Colour online) 
The pressure  versus $L$  { a) for neutron matter at $\rho=0.08$ fm$^{-3}$; b) for $\beta$-equilibrium stellar matter at the transition point}.
The full (empty) symbols are for Skyrme forces (relativistic models) and the asterisk for BHF.
} 
\label{Fig:L0_Pt}
\end{figure}

Let us now turn to the study of the transition pressure $P_t$.
In order to investigate what evolution of $P_t$ we should expect for increasing $L$,
we express the pressure in the parabolic-GLDM framework:
\be
P(\rho,y)
&=&\frac{\rho^2}{3\rho_0}\left[ L y^2 + \sum_{n\geq 2}\left(c_{{\rm IS}, n}+c_{{\rm IV}, n}y^2\right)\frac{x^{n-1}}{(n-1)!}\right] \; .\nonumber\\
\label{eq:pressure}
\ee
We can first notice that, for fixed density and asymmetry, the pressure should increase with L.
For instance, within the parabolic GLDM, the pressure of neutron matter at saturation is just 
$P_{\rm NM}(\rho_0)=P(\rho_0,1)=L\rho_0/3$. 
{
At subsaturation density and in neutron matter, there is still a correlation between $L$ and 
$P_{\rm NM}$ that we show in Fig.~\ref{Fig:L0_Pt}a), for the density $\rho$=0.08 fm$^{-3}$. {An equivalent result was obtained
in \cite{iida07} at $\rho$=0.1 fm$^{-3}$.}
In asymmetric matter corresponding to the transition position $(\rho_t,y_t)$, this simple relation 
is spoilt by two effects: the dependence of the transition position ($\rho_t$, $y_t$) on $L$ and
 the dependence of the isovector coefficients $c_{{\rm IV}, n}$ appearing
in the pressure~(\ref{eq:pressure}) on L.
We indeed remind the correlation between ($\rho_t$, $y_t=1-2Y_{p,t}$) and $L$ shown in 
Figs.~\ref{Fig:L0_rt_Ypt_Ksym}a)-b): $\rho_t$ decreases with increasing $L$ while
$y_{t}$ increases with increasing $L$.
From Eq.~(\ref{eq:pressure}), these additional correlations might change the evolution of 
the pressure $P_t$ with respect to $L$.
}
In Fig.~\ref{Fig:L0_Pt}b) we represent
$P_t$  versus $L$ calculated consistently for each of the models considered.
If we consider only the sub-group formed by the Skyrme models, 
it could be noticed a very slight decreasing correlation of $P_t$ for $L>$60~MeV.
However, considering all the models in Fig.~\ref{Fig:L0_Pt}, 
there is a large dispersion for the values of $P_t$, 
and it is not possible to define a general trend.
We can notice that the dispersion of $P_t(L)$ corresponds to an amplification of the dispersion
already present in $P_{\rm NM}(L)$ at $\rho$=0.08 fm$^{-3}$.
The dispersion in Fig.~\ref{Fig:L0_Pt}a) 
means that the pressure is quite sensitive to the GLDM coefficients others than $L$: 
thus, it is difficult to extract a clear relation between $P_t$ and $L$
which would be satisfied by 
{all the}
models.

{Let us now analyze the result shown in Fig.~\ref{Fig:L0_Pt}b).
In the parabolic-GLDM, the pressure~(\ref{eq:pressure}) depends 
clearly on various terms which are correlated to $L$. 
In the following, we analyze the contribution of these correlations
to the global relation between $P_t$ and $L$.}
The variation of the transition pressure $P_t$ with $L$ can  be decomposed into 
{two} different contributions:
{
the first one is related to the variation of $P_t$ with respect to the transition position, $(\rho_t,y_t)$, and
the second one is related to the variation of $P_t$ with respect to the isovector coefficients in 
Eq.~(\ref{eq:pressure}), $c_{{\rm IV},n}$.
We now introduce our notations:
the variation of the transition position induces two contributions:
$\frac{\partial P}{\partial \rho}\frac{\delta\rho_t}{\delta L}$ 
(hereafter named position~1), and 
$\frac{\partial P}{\partial y}\frac{\delta y_t}{\delta L}$ (named position 2).
The variation with respect to the isovector coefficients in the GLDM induce the terms
$\frac{\partial P}{\partial c_{{\rm IV},n}}\frac{\delta c_{{\rm IV},n}}{\delta L}$ 
(named coef~$n$).
All the partial derivatives are taken at the transition position $(\rho_t,y_t)$ 
obtained with the complete functional for each model.
}
The 
{variation}
of $\rho_t$, $y_t$ and $K_\mathrm{sym}$ 
{with respect to $L$}
is extracted from linear fits
in Figs.~\ref{Fig:L0_rt_Ypt_Ksym}a)-c), giving the values 
${\delta\rho_t}/{\delta L}=(-3.84\pm 0.24)\times 10^{-4}$ MeV$^{-1}$ fm$^{-3}$ and 
${\delta y_t}/{\delta L}=(6.08\pm 0.82)\times 10^{-4}$ MeV$^{-1}$, and
$\delta K_{\rm sym}/\delta L=3.07\pm0.33$.
{Notice that in doing so, we smooth out the dispersion that still exist 
between the different models (see Figs.~\ref{Fig:L0_rt_Ypt_Ksym}a)-c)).
However, anticipating the results, the dispersion is the largest for the variable $Y_{p,t}$ 
which has a very weak contribution to the global variation of $P_t$.}
We 
consider the contributions {of the terms} coef~$n$ up to $n=2$.
{ We have verified that no general  correlation appears between $L$ and 
$c_{{\rm IV},n}$ for $n>2$; furthermore, due to the fast convergence of the expansion 
(\ref{eq:pressure}) at  $\rho=\rho_t$, the 
contribution 
for $n>2$ is}
small.

\begin{figure}[t]
\begin{center}
\begin{tabular}{ccc}
\includegraphics[width =0.9\linewidth]{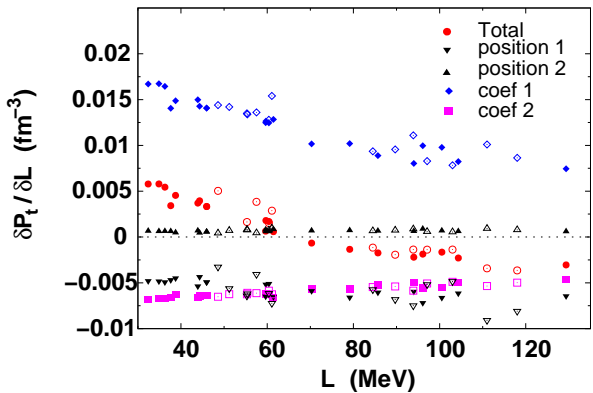}&
\end{tabular}
\end{center}
\caption{(Colour online) Expected variation of $P_t$ with $L$, comparing the different contributions:
variation due to the 
{ position, position~1 (down-triangles) and  position~2 
(up-triangles),  and due to  $c_{{\rm IV},n}$ coefficients, coef~1 (diamonds)  and coef~2 (squares).}
The resulting total pressure variation is very close to zero (circles).
The full (empty) symbols are for Skyrme forces (relativistic models).}
\label{Fig:Pt_variation}
\end{figure}

We show in Fig.~\ref{Fig:Pt_variation} the 
contributions 
{of the terms position~1, position~2, coef~1 and coef~2}
to the variation of $P_t$.
The contribution coef~1, namely 
$\frac{\partial P}{\partial L}=\rho^2 y^2 / (3\rho_0)$,
is clearly large and positive, as expected.
This term alone would predict an increase of the transition pressure with $L$, which is not observed.
However, this term is balanced by the sum of the two terms position~1 and coef~2, 
which are negative and of similar magnitudes
{(down-triangles and squares in Fig.~\ref{Fig:Pt_variation})}. 
The term position~2 brings only a negligible contribution.
The sum of the different contributions, represented by the circles in Fig.~\ref{Fig:Pt_variation}, 
is thus very close to zero; furthermore, its sign depends on the model.
We conclude that there is no clear correlation between $L$ and $P_t$,
due to the non-trivial balance between opposite contributions which differ among models. 
This can explain why  opposite behaviors  have been obtained for  $P_t(L)$ in the literature~\cite{Xu09,Moustakidis}.

It is important to notice the role of the $L-K_{\rm sym}$ correlation,
leading to the contribution coef~2:
\be
\frac{\partial P}{\partial K_{\rm sym}}\frac{\delta K_{\rm sym}}{\delta L}
=\frac{\rho_t^2y_t^2(\rho_t-\rho_0)}{(3\rho_0)^2}\frac{\delta K_{\rm sym}}{\delta L} \,.
\nonumber
\ee
If this term had not been considered, 
the prediction that $P_t$ increases with $L$  would have persisted, 
since the term position~1 is not sufficiently large to overcome the term coef~1.
In other words, the decrease of $\rho_t$ is not sufficient to predict a decrease of $P_t$
if we consider only the leading term of the GLDM development in Eq.~(\ref{eq:pressure}).
The correlation $L-K_{\rm sym}$ is 
{therefore} crucial in this discussion.

In the present letter we have explained why there is a good correlation between the crust-core
transition density and the symmetry energy slope, $L$; 
and we predict that this behaviour should not depend on the relation between $L$ and $K_{\rm sym}$.
On the contrary, no correlation of the transition pressure with $L$ was obtained.
We have highlightened the competing contributions to the variation of $P_t$ with $L$:
this explains that the effect of $L$ on the transition pressure is not strong enough 
to overcome the dispersion due to the interaction details of independent models.
This means that an experimental determination of $L$ alone
will not be sufficient for a good estimation of the crust mass and moment of inertia of a compact star. 
In fact, the range of variation of $P_t$ obtained in the present letter
with a large set of nuclear models lies within the interval indicated in \cite{pethick95}, $0.20<
P_t< 0.65$ MeV/fm$^3$, 
but completely out of the interval obtained in \cite{Xu09},
where a correlation between $P_t$ and $L$ was supposed
and the range of $L$ values was determided from isospin diffusion.
The large dispersion of the predicted transition pressure obtained when independent models
are considered needs to be reduced.
A more accurate knowledge of the isoscalar EOS could improve the situation.
Experimental constraints given directly at subsaturation density for the isovector EOS would also help.
The striking correlation between $K_{\rm sym}$ and $L$ appears to be an important feature,
which should be further investigated.

\vspace{1cm}

We would like to thank Isaac Vida\~na for the BHF results. 
This work was partially supported by the ANR NExEN contract, 
FCT (Portugal) under grants SFRH/BPD/46802/2008, 
FCOMP-01-0124-FEDER-008393 with FCT reference CERN/FP/109316/2009, 
and COMPSTAR, an ESF Research Networking Programme.

\end{document}